\documentstyle[aps,12pt,epsf]{revtex}
\begin{document}
\title{Modeling  Pionic Fusion}
\author{Alexander Volya, Scott Pratt and Vladimir Zelevinsky}
\address{National Superconducting Cyclotron Laboratory,
Michigan State University,\\
East Lansing, Michigan 48824-1321, USA}
\date{\today}
\maketitle

\begin{abstract}
\baselineskip 14pt

Recently observed rare heavy ion fusion processes, where the entire available
energy is carried away by a single pion, is an example of extreme collectivity
in nuclear reactions.  We calculate the cross section in the approximation of
sudden overlap, modeling the initial and final nuclei by moving harmonic
oscillator potentials. This allows for a fully quantum-mechanical treatment,
exact conservation of linear and angular momenta and fulfillment of the Pauli
principle. The results are in  satisfactory agreement with data. Mass
number dependence and general trends of the process are discussed.

\end{abstract}

\baselineskip 14pt
\section{Introduction}
Nuclear fusion reactions which produce a pion are often referred to as pionic
fusion.  Pion production has been observed \cite{horn96,bornec81,waters93} at
energies approaching absolute threshold, where the entire available energy is
converted into the pion, demonstrating an amazing collective behavior of
nucleon systems. However it remains quite difficult to incorporate the observed
collectivity into existing theoretical models. A variety of studies
\cite{miller87,erazmus91,braun84,suzuki} have dealt with subthreshold pion
production in heavy ion collisions, where the energy per nucleon is below the
energy threshold of the elementary single-particle reaction $N N\rightarrow
NN+\pi$. Most models, such as those featuring pion bremsstrahlung mechanisms
\cite{vasak80,vasak85}, quantum molecular dynamics approaches\cite{li91},
perturbative calculations using Boltzmann-Nordheim-Vlasov equations
\cite{bonasera1990} or nuclear structure functions \cite {providencia1988},
provide a good picture at energies starting from $E/A\approx 30$ MeV
up to the
single-particle threshold $E/A=280$ MeV in the laboratory frame.  In the
present work motivated by the experimental results of
\cite{horn96,bornec81,waters93} our aim is to consider even lower energies and
study the behavior of the cross section of fusion reaction in the region down
to $\approx$10 MeV above the absolute threshold.  This necessitates a careful
consideration of limitations on the reaction given by conservation laws and the
Pauli exclusion principle which govern the behavior of the cross section in
this extreme situation. The statistical approach used in most existing models
at higher energy has to be substituted by low energy many-body structure
physics.

\begin{figure}
\begin{center}
\epsfxsize=8.0cm \epsfbox{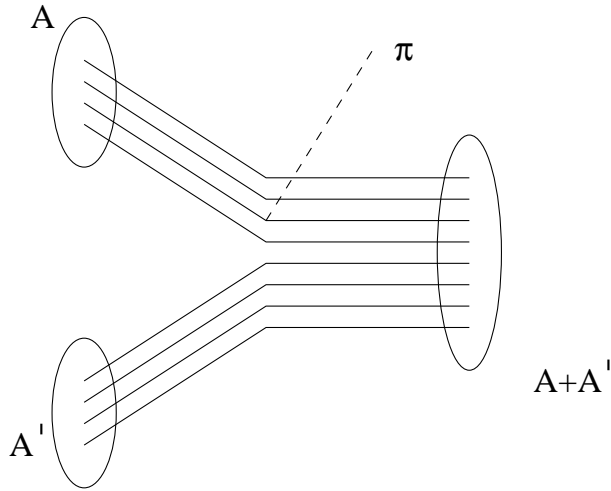}
\end{center}
\caption{The pionic fusion of two nuclei in the sudden approximation
is illustrated.
\label{picture}}
\end{figure}

Our model, that is described in the Sect. \ref{sec_2}, considers the cross
section in the Born approximation,  assuming that pion production occurs
through coupling to a single nucleon.  All possible further rescatterings of the
pion are
expected to significantly reduce the probability of the reaction,
and  are
ignored as   higher order processes. A schematic picture of the
reaction is shown in Fig. \ref{picture} demonstrating the pionic fusion of two
nuclei $A$ and $A^{\prime}$.
Many-body nuclear mean field parameters are assumed to be constant and suddenly
change from the initial to the final values. This will be referred to as the
sudden approximation.  The three-dimensional harmonic oscillator shell model is
used to describe the structure of the incoming and outgoing nuclei.  This
allows  analytical calculation of all necessary overlap amplitudes.  The
stationary wave functions are constructed as Slater determinants projected onto
good angular momentum. Taking into account the center-of-mass motion we preserve
linear momentum.  Sect. \ref{sec_3} shows the implementation of the model for
the case of pionic fusion of two identical nuclei. In Sect. \ref{sec_4} we
present a low pion momentum approximation, for which more general results could
be derived.  The parallel discussion of mathematical details is given in the
Appendices.  The application of this model to experimentally observed pionic
fusion reactions shows a good agreement with data. The comparison is presented
in Sect.  \ref{sec_5} along with some predictions  for heavy
nuclei.

\section{Description of the  model  }
\label{sec_2}
\subsection{The transition amplitude}
We approach the problem of pionic fusion as a stationary scattering problem. We
consider the reaction cross section to be given by the Fermi golden rule in
terms of the transition amplitude $\langle F|H|I\rangle$, where $I$ and $F$
refer to the initial and final states, respectively, of the whole system
including the emitted pion.  The density of final pion states is given by $ V
k^2 d k\, d\Omega/(2\pi)^3$ with $k$ as a momentum of the pion produced;
$\Omega$ is a solid angle in the center of mass (CM) frame, and $V$ stands for
the quantization volume.  In all further calculations the pion is assumed to be
fully relativistic whereas nucleons obey non-relativistic quantum mechanics. We
use a set of natural units with $\hbar=c=1$.  In this framework the
differential cross section can be written as
\begin{equation}
d\sigma={\omega k m \over 2p_n (2\pi)^2} |\langle F|H|I\rangle |^2 V^2 d\Omega \quad ,
\label{dsigma}
\end{equation}
where $m$ is a nucleon mass,  $p_n$ is the CM momentum per nucleon in
the initial state and $\omega=\sqrt{k^2+m_{\pi}^2}$ is  pion energy.

On the single-nucleon level one can use a  phenomenological Hamiltonian
density for the pion-nucleon interaction \cite{koltun1966},
\begin{equation}
{\cal H}=g{\overline{\psi}} {\gamma_5} {\vec \tau}\psi{\vec
\pi}+4\pi{\lambda_1\over m_\pi} \bar{\psi}{\vec\pi}\cdot {\vec \pi}\psi+
4\pi{\lambda_2\over m_\pi^2} \bar{\psi}{\bf r}\cdot{\vec \pi}\times
\dot{\vec \pi} \psi \quad .
\label{interaction2}
\end{equation}
A number of studies have been performed analyzing this form of the interaction
within the context of chiral perturbation theory \cite{sato97}.  The first
term, often called in the literature the impulse or Born term, is responsible
for single-pion production in a $p$-wave. We neglect the second and the third
$s$-wave terms which require an additional interaction to absorb the extra pion
created in the four-point vertex. We believe that
due to the  difficulty of recombining the  nucleons into an
appropriate final state the second and third terms become increasingly
unimportant for larger nuclei.
It has also been experimentally observed
that in the pionic fusion reactions the pion is predominantly produced in the
$p$-wave \cite{horn96,bornec81}.  Reduction of the first term in the
Hamiltonian to a non-relativistic case gives an interaction of the form
\begin{equation}
\Gamma=g{{\vec\sigma}\cdot{\bf{k}}\over 2m}\quad ,
\end{equation}
with the coupling $g$ appropriately defined according to isospin.
Separation of  the quantized pion field,
\begin{equation}
\pi(x)=\sum_{k} {1\over{\sqrt{2\omega V}}} ( a_{k}^{+} e^{-i{\bf k}
\cdot {\bf x}} +
a_{k} e^{+i{\bf k} \cdot {\bf x}}),
\end{equation}
in the matrix element of Eq. (\ref{dsigma}) reduces the transition amplitude to
the following form
\begin{equation}
\langle F|H|I\rangle = {1\over{\sqrt{2\omega V}}} {1\over 2m}
\, \langle f|\sum_{{\rm  nucleons}}g\, {\bf k} \cdot{\vec \sigma}
e^{-i{\bf k} \cdot {\bf x}}|i\rangle  \quad ,
\label{eikx}
\end{equation}
where $|f\rangle $ and $|i\rangle $ are final and initial states
of the nucleon system.
\subsection{Nuclear wave functions}
We will approximate a state of a nuclear system with an antisymmetric
combination built upon single-particle (s.p.)  states. We take these
states  from the harmonic
oscillator shell model, which allows for the analytic calculation of
corresponding overlaps. The approach however can be extended to any
single-particle basis.  Each of the single-particle states can be
characterized by the number of excitation quanta in three Cartesian directions,
the nucleon spin and isospin projections. The locations of the centers of the
harmonic oscillator potentials for all separate nuclei have to be introduced as
additional parameters to the wave function. The importance of these parameters
in projecting a nucleon wave function onto a state with correct total momentum
for every nucleus participating in the process is discussed below.  Following
these assumptions we will write the wave function of a nucleon system as
follows
\begin{equation}
 |\underbrace{({\vec \alpha}_1,s_1,t_1;{\vec \alpha}_2,s_2,t_2;\ldots;{\bf
r})}_{{\rm nucleus}\,A},\underbrace{( \overbrace{{\vec
\alpha}_{A+1},s_{A+1},t_{A+1}}^{(A+1){\rm th \,\,s.p.\, \,state}};{\vec
\alpha}_{A+2},s_{A+2},t_{A+2};\ldots;{\bf r}^{\,\prime})}_{{\rm
nucleus}\,A^{\prime}} \rangle \quad .
\end{equation} 
In this example we assume that the system consists of two nuclei $A$ and
$A^{\prime}$ with the centers of their respective harmonic oscillator potentials at
${\bf r}$ and ${\bf r^{\,\prime}}$. The single-particle orbitals are numbered
from $1$ to $A$ for the first nucleus and from $A+1$ up to the total number
of nucleons $A_f=A+A^{\prime}$ for the second one. Labels ${\vec
\alpha}=(\alpha_x,\alpha_y,\alpha_z)$ are Cartesian quantum numbers of
single-particle states, while $s$ and $t$ are the spin and isospin projections.
Protons and neutrons can be considered separately as well as different spin
projections of the nucleons, reducing the wave function of the state to a
product of four components.  If the described separation is performed and the
resulting part of the wave function contains only single-particle states with
the same values of either $s$ or $t$ then the corresponding index is omitted in
writing.  We use a standard form for the one-dimensional harmonic
oscillator wave functions centered at $r$ in
the coordinate representation:
\begin{equation}
\langle x|(\alpha;r)_v\rangle =\sqrt{v \over \sqrt{\pi} 2^\alpha {\alpha}!}\,
H_\alpha(v(x-r)) e^{-v^2 (x-r)^2/2} \quad .
\label{harmonic}
\end{equation}
The parameter $v$ is defined for a single oscillator as
$v=\sqrt{m\omega}$. These parameters characterize the mean
field potentials for every incoming or outgoing nucleus.
The function $H_n(x)$ is the $n$th order Hermite polynomial of the variable $x$.
The discussion of the overlap integrals  such as 
$\langle (\alpha^{\prime},r^{\,\prime})_{v^{\prime}}|(\alpha,r)_v\rangle $, and
the general form of the results is presented in Appendix A.

A simple projecting technique  was used to construct 
wave functions  as  eigenstates of the momentum
operators that correspond to the total
momenta of each individual  nucleus,
\begin{eqnarray}
\nonumber
|({\vec
\alpha}_1,s_1,t_1;\ldots;{\bf p}),
({\vec \alpha}_{A+1},s_{A+1},t_{A+1};\ldots;{\bf p}^{\,\prime})\rangle = \\ {\cal N}^{-1}\,
\int\int_{-\infty}^{+\infty} d^3 r\, d^3 r^{\,\prime}\, |({\vec
\alpha}_1,s_1,t_1;\ldots;{\bf r}),
({\vec \alpha}_{A+1},s_{A+1},t_{A+1};\ldots;{\bf r}^{\,\prime})\rangle  e^{i({\bf p} \cdot
{\bf r}+{\bf p}^{\,\prime} \cdot {\bf r}^{\,\prime})} \quad .
\label{momentum}
\end{eqnarray}
It is easy to check that
\begin{equation}
-i\sum_{j=1}^{A}{\bf \nabla}_j |({\vec
\alpha}_1,\ldots;{\bf p}),
({\vec \alpha}_{A+1},\ldots;{\bf p}^{\,\prime})\rangle = {\bf p}\,|({\vec
\alpha}_1,\ldots;{\bf p}),
({\vec \alpha}_{A+1},\ldots;{\bf p}^{\,\prime})\rangle 
\end{equation}
and
\begin{equation}
-i\sum_{j=A+1}^{A+A^{\prime}}{\bf \nabla}_j |({\vec
\alpha}_1,\ldots;{\bf p}),
({\vec \alpha}_{A+1},\ldots;{\bf p}^{\,\prime})\rangle = {\bf p}^{\,\prime}\,|({\vec
\alpha}_1,\ldots;{\bf p}),
({\vec \alpha}_{A+1},\ldots;{\bf p}^{\,\prime})\rangle \quad . 
\end{equation}
In the above example the situation with two-nuclei state is shown, which is
appropriate for describing the initial state in pionic fusion. The final state
containing just one fused nucleus is constructed analogously.

Due to the finite range of the interaction, the overall normalization ${\cal
N}$ of the state (\ref{momentum}) that contains several moving nuclei, is just
a product of normalizations for each of the constituent nuclei individually. It
is useful to write the CM coordinates separately from the relative coordinates
of the nucleons
\begin{equation}
|({\vec \alpha}_1;{\vec \alpha}_2; \ldots {\vec \alpha}_A ;{\bf r})_v\rangle =
|({\vec \alpha}_{\rm CM}=(0,0,0);{\bf r})_{v\sqrt{A}}\rangle\, |\psi_{\rm
rel}\rangle \quad .
\end{equation}
The relative coordinate wave function $|\psi_{\rm rel}\rangle $ can be
complicated but the CM part for the ground state nucleus is simply represented
by the unphysical ground state oscillation of the center of mass in the
effective harmonic potential with the parameter $v\sqrt{A} $. This is removed
by a projection (\ref{momentum}) onto the correct momentum state.  The
normalization integral can be expressed as 
\begin{eqnarray}
\nonumber
{\cal N}^2=\int\int d^3r\, d^3r^{\,\prime}\langle ({\vec \alpha}_{\rm
CM};{\bf r}^{\,\prime})_{v\sqrt{A}}|({\vec \alpha}_{\rm CM};{\bf r})_{v\sqrt{A}}\rangle \,\langle \psi_{\rm
rel}|\psi_{\rm rel}\rangle  e^{i {\bf p}
({\bf r}-{\bf r}^{\,\prime})}\\
=\int d^3r \int d^3r^{\,\prime}  e^{-A({\bf r}-
{\bf r}^{\,\prime})^2 v^2/4} e^{i {\bf p}
({\bf r}-{\bf r}^{\,\prime})}=\left ({4\pi \over v^2 A}\right )^{3/2} V e^{-{p^2
/ A v^2}} \quad . 
\label{Q}
\end{eqnarray}
A different method of calculating the normalization along with further
justification of this form for the CM part of the wave function is discussed
in Appendix B.
We also note here that with a slight modification of Eq. (\ref{Q}) the
orthogonality of the nucleon wave functions can be shown
$$
\langle({\vec \alpha}_1;{\vec \alpha}_2; \ldots {\vec \alpha}_A ;{\bf p}^{\prime})_{v}|({\vec \alpha}_1;{\vec \alpha}_2; \ldots {\vec \alpha}_A ;{\bf p})_v\rangle
={\cal N}^2 \delta_{{\bf p},\,{\bf p}^{\prime}} \, .
$$
\section{Fusion reactions $A+A \rightarrow 2A +\pi$}
\label{sec_3}
For the remainder of the paper we will assume $A$ to be the mass number of each
of the initial nuclei with proton-neutron composition ($Z$,$N$) and $w$ the
oscillator parameter. The entire initial state is characterized by a set of the
single-particle quantum numbers $\{{\vec \alpha}_i \}$. The fusion product has
$2A=A_f$ nucleons, the oscillator parameter $v$, and the final state quantum
numbers $\{{\vec \beta}_i \}$. The collision is considered in the CM reference
frame; therefore we use ${\bf p}$ and $-{\bf p}$ to denote the momenta of the
incoming nuclei and ${\bf k}$ for a final pion momentum with corresponding
${\bf p}_f=-{\bf k}$ as the total momentum of the recoil nucleus. The
integration of the wave functions leading to correct momenta, Eq. (\ref{momentum}), is performed at a final stage so initially overlaps are calculated as functions of ${\bf r}$ , ${\bf r}^{\,\prime}$ and ${\bf R}$, the locations of the
centers of the two initial nuclei and the final nucleus, respectively. 

\subsection{Charged pion production}
\begin{figure}
\begin{center}
\epsfxsize=8.0cm \epsfbox{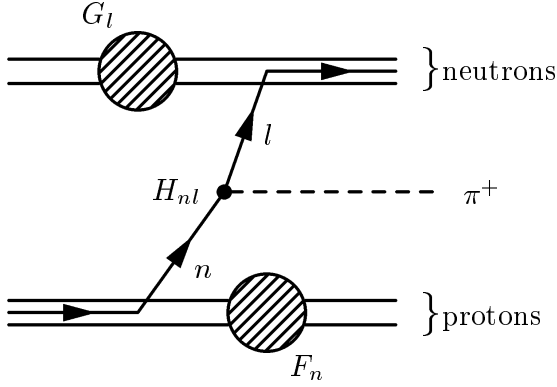}
\end{center}
\caption{One of the amplitudes of the total fusion process: an initial proton
from the $n$th orbit produces a $\pi^{+}$ and ends at the $l$th final neutron
single-particle orbit. $F_n$ is the remaining overlap of a proton system with
the $n$th initial single-particle state missing. $G_l$ is the neutron
overlap with no $l$th state in the final system.
\label{oneampl}}
\end{figure}

We begin with the case of $\pi^{+}$ production where one of the initial protons
interacts with the pion field producing a neutron and a real on-shell pion.
With the assumption that the pion was produced in a single-particle
interaction, the total amplitude of the process becomes a sum over all possible
amplitudes shown in Fig. \ref{oneampl}, with the pion vertex connecting any of
the initial protons to any of the final state neutrons with the correct
relative sign to preserve antisymmetry.
Suppose the interacting proton in the single-particle state $n$ produced a
neutron in the state $l$ of the final nucleus. In the initial state we sum over
the occupied orbitals of the first and the second colliding nucleus, for $n
\leq Z$ and for $Z<n\leq 2Z$, respectively. We use the notations $G_l
(r,r^{\prime},R)$ for the neutron overlap
\begin{equation}
G_l=\langle \underbrace{({\vec \beta}_1,\tilde{s}_1;\ldots {\vec
\beta}_{l-1},\tilde{s}_{l-1}; {\vec
\beta}_{l+1},\tilde{s}_{l+1};\ldots ; {\bf R})}_{{\rm no\,} l {\rm
th\, s.p.\,state}}|({\vec\alpha}_1,s_1;\ldots ;{\bf
r})({\vec\alpha}_{A+1},s_{A+1};\ldots ; {\bf r^{\,\prime}})\rangle \, , 
\end{equation}
$F_n(r,r^\prime,R)$ for the proton overlap
\begin{equation}
F_n=\langle ({\vec \beta}_1,\tilde{s}_1;\ldots ;{\bf
R})|\underbrace{({\vec\alpha}_1,s_1;\ldots;{\vec\alpha}_{n-1},s_{n-1};{\vec\alpha}_{n+1},
s_{n+1};\ldots)}_{{\rm no\,} n {\rm th\, s.p.\,state}}\rangle  \, ,
\end{equation}
and $H_{n\,l}$ for a single-particle matrix element
\begin{equation}
H_{n\,l}=\left \{ \matrix {\langle ({\vec \beta}_l,\tilde{s}_l;{\bf R})|g\,{\vec \sigma}
\cdot {\bf k} \, e^{-i{\bf k}\cdot{\bf x}}|({\vec
\alpha}_n,s_n; {\bf r})\rangle  \qquad n \leq Z \cr
\langle ({\vec \beta}_l,\tilde{s}_l;{\bf R})|g\,{\vec \sigma}
\cdot {\bf k} \, e^{-i{\bf k}\cdot {\bf x}}|({\vec
\alpha}_n,s_n;{\bf r}^{\,\prime})\rangle  \qquad n >  Z \cr} \right
. \, .
\end{equation}
Finally, following Eq.(\ref{eikx}), the total amplitude
can be  expressed in terms of the  following sum:
\begin{eqnarray}
\langle F|H|I\rangle = {1\over {\cal N}_i {\cal N}_f}{1\over{\sqrt{2\omega V}}} {1\over 2m}
\int \int d^3r\, d^3r^{\prime}\, d^3R \,\sum_{n l}(-1)^{n+l} F_n G_l H_{n
l} e^{i{\bf p}({\bf r}-{\bf r}^{\,\prime})-i{\bf p}_f\cdot {\bf R}} \,
.
\label{sum}
\end{eqnarray}

The
determinants of the matrices are constructed from a  product of the
single-particle
overlaps of  size  
$(2N)\times (2N) $ for the neutrons ($G_l$) and $(2Z-1) \times (2Z-1)$ for
the protons ($F_n$). The Gaussian nature of the single-particle
overlaps  allows one to separate all
exponential factors that govern the general trend of the cross section
leaving only some polynomials of a general form that carry 
spin, isospin and Pauli blocking information. These 
mathematical manipulations are discussed in some detail in Appendix B.
Here we  present a final expression for the square of the transition amplitude
\begin{equation}
|\langle F|H|I\rangle |^2={2 g^2\over  V^2}\left ({2\pi\over
v^2 A}\right )^{3/2}
{|{\bf k}|^2\, \xi \over \omega
m^2}\, \eta^{6(A-1)}
\left |M_{+}
\right |^2\, ,
\label{ppamp}
\end{equation}
in which the exponential factor $\xi$,  effective oscillator parameter $\eta$ and
reduced amplitude $M_{+}$ 
are introduced as follows
\begin{equation}
\xi \equiv \exp{\left (-{2p^2\over Av^2}-{k^2\over v^2+w^2}-
{k^2(w^2-v^2)\over 2Av^2(v^2+w^2)}\right )} \quad ,
\end{equation}
\begin{equation}
\eta \equiv {2vw\over v^2+w^2}\quad ,
\label{eta}
\end{equation}
and
\begin{equation}
M_{+} \equiv P(k,p)e^{{{\bf p} \cdot {\bf k}}/A v^2} +
Q(k,p)e^{-{{\bf p} \cdot {\bf k}}/A
v^2} \, .
\label{M}
\end{equation}
Here, $P$ and $Q$ are dimensionless polynomials of $p$ and $k$, the total CM
momentum of the initial nuclei and the final pion momentum. The polynomials are
to be determined using particular configurations of the initial and final
nuclei. They are also functions of $v$ and $w$ which determine the appropriate
momentum scale.  If the two colliding nuclei have the same initial shell model
state then $P(k,p)=\pm Q(k,-p)$ (the phase difference given by $\pm$
sign for even or odd $Z$, respectively, is due to imposed Pauli
antisymmetry, see
Appendix B).  The procedure of analytically calculating $P$ and $Q$ involves
finding the determinant of the matrix constructed from polynomials that result
from integrating a product of Hermite polynomials of the form $\langle
(\beta,R)|(\alpha,r)\rangle $; and performing the integrational Fourier-type conversion,
Eqs. (\ref{momentum}).  This process is discussed in Appendix B. The size of
the matrices is determined by the number of nucleons of the same spin-isospin
type.

\subsection{Neutral pion production}
The case of $\pi ^{0}$ production can be considered in a similar fashion. A
neutral pion can be produced either by one of the protons or by one of the
neutrons, which couple with a negative relative sign.  Compared to charged
pions the coupling is larger by a factor $\sqrt{2}$.  The final amplitude can
then be expressed, similarly to Eq. (\ref{ppamp}), as
\begin{equation}
|\langle F|H|I\rangle |^2={g^2\over  V^2}\left ({2\pi\over
A\,v^2}\right )^{3/2}
{|{\bf k}|^2\,\xi\over \omega
m^2}\,\eta^{6(A-1)} 
\left | M_{0}
\right |^2 \, .
\label{p0amp}
\end{equation}
Here the reduced amplitude can  be split into a proton and a
neutron part:
\begin{equation}
M_{0}=P_p(k,p)e^{{\bf p \cdot k}/A v^2} + Q_p(k,p)e^{-{\bf p \cdot k}/A
v^2}- P_n(k,p)e^{{\bf p \cdot k}/A v^2} - Q_n(k,p)e^{-{\bf p \cdot k}/A
v^2} \, .
\label{p0mat}
\end{equation}
\section{Low pion momentum approximation}
\label{sec_4}
Due to the  specific form of the polynomials discussed above,  further
simplifications can be made for the case of  $\pi^{0}$
production. Near the absolute threshold, 
the pion momentum $|{\bf k}|$ is small compared to all  other
momentum parameters
$|{\bf p}|$, $v$ and $w$, and can be ignored in polynomials. Then
\begin{eqnarray}
\nonumber
H_{n\,l}=\langle ({\vec \beta}_l,s_l;{\bf R})_v|{\vec \sigma} \cdot
{\bf k}\,e^{-i{\bf k} \cdot {\bf x}}|({\vec\alpha}_n,s_n;{\bf
r})_w\rangle
\\
\approx 
\langle s_l|{\vec \sigma} \cdot {\bf k}|s_n\rangle \,\langle ({\vec \beta}_l;{\bf
R})_v|({\vec\alpha}_n; {\bf r})_w\rangle \exp \left ({\frac{-k^2}
{2(v^2+w^2)}}-{\frac{i{\bf k}\cdot ({\bf R}v^2+{\bf
r}w^2)}{(v^2+w^2)}}\right ) \, .
\label{hnl}
\end{eqnarray}
With this approximation, the interaction part is factorized into
exponents as shown in the expression above. Therefore  the total pionic
fusion amplitude is a
product of a pure fusion  amplitude and the expression
that arises from the operator ${\vec \sigma} \cdot {\bf k}$ acting
on the nucleons.
For a  given type of the 
initial and final nucleon,
the sum of a single-particle matrix element
multiplied by  the corresponding overlap of the remaining particles reduces to a
sum of matrix elements multiplied by  the corresponding minor which is related
to a determinant of a full matrix.
It is shown in Appendix C that the polynomials can be expressed in an
analytical form if  all inner
harmonic oscillator shells are completely filled without any
gaps in all participating  nuclei. This restriction allows any type of particle-hole excitations
within the outer unfilled shell.

The total differential cross section for a neutral pion production close to 
absolute threshold is given  in the form:
\begin{eqnarray}
\nonumber
{d\sigma\over d\Omega}={g^2 A k^3 \over (2\pi)^2 2 p
m} \left (
{2\pi\over A\,v^2}\right )^{3/2}\,
\eta^{6(A-1)+{\cal Q}_f+{\cal Q}_i}
\left ({4w\over Av}\right )^{{\cal Q}_f-{\cal Q}_i}
e^{-{2p^2/ Av^2}}
\times
\\
\left |\,2^{q_z/ 2} T_{q_z}\left (
{ip\sqrt{2\over A \eta v w }} \right ) {(q_x-1)!! \:
(q_y-1)!!}\, \right |^2\, {1 \over \gamma^2} \,\left 
|{\tilde M}\right |^2
\quad .
\label{generic}
\end{eqnarray}
Here the integers $q_j,\,j=x,y,z$  are introduced as differences between
numbers of quanta in final and initial systems for three Cartesian directions;
${\cal Q}_i$ and ${\cal Q}_f$ are total numbers of quanta in initial and final
systems.  These values are defined as
\begin{equation}
q_j=\sum_{\rm nucleons} {\beta}_j - \sum_{\rm nucleons}
{\alpha}_j\, , \quad {\cal Q}_i=\sum_{\rm nucleons} (\alpha_x+\alpha_y+\alpha_z)\,
, \quad
{\cal Q}_f=\sum_{\rm nucleons} (\beta_x+\beta_y+\beta_z)\, .
\end{equation}
The spin and radial parts of the wave function are completely decoupled
in our non-relativistic description of the nucleon system. This allows to
introduce the matrix element used in Eq. (\ref{generic})
\begin{equation}
{\tilde M}={1\over |{\bf k}|}\,\langle {\tilde f}|\sum_{\rm nucleons} \tau_z\,{\vec \sigma} \cdot {\bf
k}|{\tilde i}\rangle \, ,
\end{equation}
where ${\tilde i}$ and  ${\tilde f}$ are the  spin-isospin parts of nucleon wave
function of initial and final systems, respectively. This matrix
element could be directly computed  for every particular nuclear
configuration, but for a large number of states degenerate within harmonic
oscillator model it is useful to use an approximation for the average
\begin{equation}
\overline{{\tilde M}} \,=\,\overline{\left
({Z_{\uparrow}-Z_{\downarrow}-N_{\uparrow}+N_{\downarrow}}\right
)}\,  .
\label{whatever}
\end{equation}
The Cartesian directions of the harmonic oscillator quantization axes are
chosen in such a way that the $z$ axis coincides with the beam direction,
though the spin is quantized along the ${\bf k}$ axis that simplifies the
action of ${\vec \sigma} \cdot {\bf k}$ which is used to obtain
Eq. (\ref{whatever}). 
Integers $Z_{\uparrow}$, $Z_{\downarrow}$, $N_{\uparrow}$ and $N_{\downarrow}$
are mean numbers of particles for each spin-isospin combination with respect to our axis
of spin quantization. The polynomials $T_n(x)$, defined in Eq. (\ref{T_n}) of
Appendix A, can be approximated as
\begin{equation}
2^{q_z/ 2} T_{q_z}\left (
{ip\sqrt{2\over A\eta v w}} \right ) \, \approx \,\left (
{ip\sqrt{2\over A \eta v w}} \right )^{q_z} \quad . 
\end{equation}
This approximation is valid in the limit that the arguments become
large and allows for a better quantitative understanding of the
behavior of the cross section. The value of the argument is almost
independent of the mass number $A$ at  threshold energy: 
$$p\sqrt{2\over A \eta v w}\,\approx \, 6 \, .$$

In Eq. (\ref{generic}) only the lowest order term in the pion momentum is
retained resulting in a $p$-wave cross section (exponents with $k$ are also
ignored).  The equation has only one numerical parameter $\gamma$, the origin
of which is discussed in Appendix C. This parameter is a product of four
factors, one for every spin/isospin nucleon species. Each factor depends on
the number of particles of corresponding type and on their distribution within
the highest harmonic oscillator shell for both initial and final nuclei.
Numerically, $\gamma$ range from $1$ to $10$ for light nuclei.  The cross
section can be zero if some symmetries are not preserved (spin, isospin,
oscillator symmetry) as well as by virtue of Eq. (\ref{oddeven}) in Appendix A
if creation of the final system requires an odd number of quanta relative to
the initial system in any of the transverse directions.

\section{Application of the model and results}
\label{sec_5}
\subsection{The reaction $p+p \rightarrow d+\pi^{+}$}
The first and the simplest example to calculate
is the two-nucleon fusion reaction
$p+p \rightarrow d+\pi^{+}$. This example serves here only for illustrative
purpose as we do not include pion rescattering due to the full interaction
given by Hamiltonian density of Eq. (\ref{interaction2}) which is important for
this elementary process.  Moreover, the deuteron hardly can be approximated
with the harmonic oscillator shell model. The polynomials $P$ and $Q$ in this
case do not depend on $p$ being equal to the matrix element of ${\vec
\sigma}\cdot {\bf k}/|{\bf k}| $ evaluated between the spinors of initial
interacting proton and final neutron. In Eq. (\ref{ppamp}) we choose a minus
sign for antisymmetry. Here, $P$ and $Q$ correspond to the choice of the first
or second initial proton to produce a pion, respectively.

Dominant partial wave channels are summarized in the following table along with
our results for their reduced amplitudes. The table was constructed by
separation of initial singlet and triplet states of the $N N$ system.  Partial
waves of the $\pi-d$ system printed in the left column that yield the dominant
contributions to the amplitudes which are shown in the right column.
\begin{equation}
\begin{array}{ccc}
{\rm pion} & NN \, {\rm state} & {\rm amplitude} \\
\hline
s{\rm -wave} & ^3P_1 & 2\,{\sqrt{2}}\,\sinh ({\frac{k\,p\,\cos\theta}{A\,{v^2}}}) \\
\begin{array}{c}
p{\rm -wave} \\ p{\rm -wave}
\end{array} &
\left . \begin{array}{c}
^1S_0 \\ ^1D_2 
\end{array}  \right \}&
2\,\cosh ({\frac{k\,p\,\cos\theta}{A\,{v^2}}})
\end{array}
\end{equation}
As can be seen from the table above, this cross section is predominantly
$p$-wave in nature at low pion energies.  The $s$-wave contribution that comes
\cite{koltun1966} from rescattering of the pion due to the interaction
(\ref{interaction2}) was not included.  The total cross section averaged over
spin projections in the initial state and summed over final states is
\begin{equation}
{d\sigma \over d \Omega}= {\frac{{g^2}\,k^3}{2\,m\,p\,{\sqrt{2\,\pi }}\,{v^3}}}
\,{\exp\left(-{\frac{4\,{p^2}+ k^2}{2\,{v^2}}}\right )}
\left[ 3\,\cosh \left ({\frac{2\,k\,p\,\cos\theta}{{v^2}}}\right ) -1
\right] \quad .
\end{equation}
The
obtained $p$-wave cross section  behaves at low energies as
\begin{equation}
 \sigma(p p \rightarrow d\pi^{+})=\tilde \sigma \left (k/
m_{\pi} \right )^3 \quad ,
\label{deuteron}
\end{equation}
where
\begin{equation}
\tilde \sigma={2\sqrt{2\pi} g^2 m_{\pi}^{5/2} \over  m^{3/2}
v^3} e^{-2 m m_{\pi}/v^2} \quad .
\end{equation}
Choosing the oscillator parameter $v=216$ MeV/c reproduces the experimental
value\cite{ericson}, $4 \tilde \sigma\approx 0.42$ fm$^2$. For this case the fusion is
sensitive to the tail of the wave function in momentum space. Since the wave
function of a deuteron is extremely non-Gaussian with a long tail in coordinate
space, choosing $v$ to reproduce the deuteron's r.m.s. charge radius would
result in a grossly underpredicted cross section. For the fusion of heavier
ions, the incoming nuclei are moving at a slower velocity and their
momenta per nucleon are similar to 
characteristic  momentum
scales of the wave functions.

The oscillator parameter $v$ can be best obtained by matching used here
harmonic oscillator type deuteron wave function to its experimentally known
behavior \cite{ericson}.  The choice of this parameter between 180 and 220 fm
would lead to the values of $4\tilde \sigma$ in the range of 0.06 to 0.48 fm$^2$. 

\subsection{The reaction $^3$He $+$ $^3$He $\rightarrow$ $^6$Li $+$ $\pi^{+}$}
\begin{figure}
\begin{center}
\epsfxsize=8.0cm \epsfbox{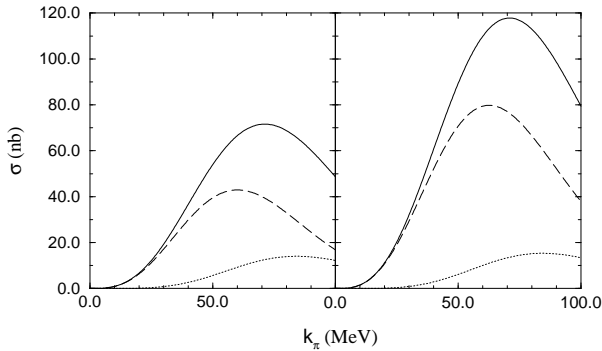}
\end{center}
\caption{Reaction cross sections for $^3$He $+$ $^3$He $\rightarrow$ $^6$Li $+$
$\pi^{+}$. The left panel shows the transition to the ground
state and the right panel to the first excited state of $^6$Li at $2.18$
MeV. The solid lines represent the total cross section,
dashed and dotted lines are
$s$ and $p$-waves, respectively.}
\label{wave}
\end{figure}

\begin{figure}
\begin{center}
\epsfxsize=8.0cm \epsfbox{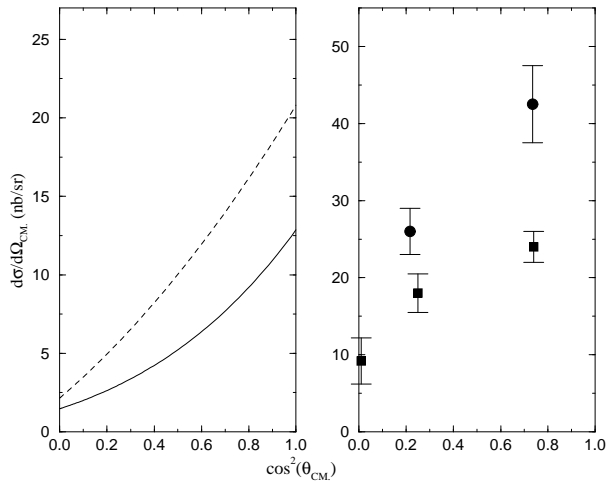}
\end{center}
\caption{Differential cross section of the reaction $^3$He $+$ $^3$He
$\rightarrow$ $^6$Li $+$ $\pi^+$. On the left panel the solid line represents
the transition to the ground state of $^6$Li and the dashed line to the first
excited state; the corresponding absolute values of pion momentum are 96 and 90
MeV/c, respectively.  The right panel displays the experimentally observed
values [2] of the differential cross section of the transition to
the ground state (squares) and to the first excited state (circles) of $^6$Li.}
\label{Li6s}
\end{figure}

As a next step, we apply the model to the experimentally studied pionic fusion
reaction $^3$He $+$ $^3$He $\rightarrow$ $^6$Li $+$ $\pi^{+}$, where even first
excited states of the $^6$Li nucleus have been resolved \cite{bornec81}. This
reaction involves heavier nuclei so that the process of pion rescattering
becomes less important as discussed above.  The polynomials $P$ and $Q$ for
Eq. (\ref{ppamp}) can be constructed in a direct way considering the shell
model structure of all nuclei involved in the reaction. The ground $1^{+}$ and
first excited $3^{+}$ states of $^6$Li were constructed within the $p_{3/2}$
j-subshell. In Fig. \ref{wave}, the total cross section for this reaction is
calculated for the fusion into the ground state (left panel) and the first
excited state (right panel).  The contributions of the $s$-wave and $p$-wave to
the cross section are plotted together. We choose a value $v=118.91$ MeV/c for
$^6$Li as it corresponds to the oscillator frequency of 15.06 MeV, the
parameter of MK3W model \cite{warburton}.  The initial parameter $w=112.7$
MeV/c was chosen by assuming the r.m.s. size 2.14 fm of $^3$He.  In
Fig. \ref{Li6s} we show the differential cross section for this fusion reaction
going into the ground state of $^6$Li (solid line) and the first excited state
(dashed line). The beam energy is assumed to be fixed so that the corresponding
absolute values of the pion momentum are $96$ and $90$ MeV/c, respectively.

Comparison with the experiment \cite{bornec81} in which pionic fusion resolves the
few lowest levels of $^6$Li shows that we obtain a reasonable ratio of the
cross sections. However, we underpredict the magnitude by approximately $40\%$,
compared to the estimated experimental value of $111\pm 11$ nb for the ground
state transition.  We note that the result is sensitive to parameters of the
shell model and their choice in the harmonic oscillator approximation
is quite uncertain for light nuclei. For example,
a variation of the final oscillator frequency within $10\%$ range of the used
value would lead to the values of the cross section between about $20$ and
$140$ nb.  Using more realistic non-Gaussian wave functions might significantly
improve the model. We might also be underestimating the cross section due to
inherent limitations of the approach. For instance, we do not consider a
gradual change of the nuclear mean field in the process of fusion substituting
it with the sudden approximation.

\subsection {The reaction $^{12}$C + $^{12}$C $\rightarrow$ $^{24}$Mg + $\pi^0$}
\begin{figure}
\begin{center}
\epsfxsize=8.0cm \epsfbox{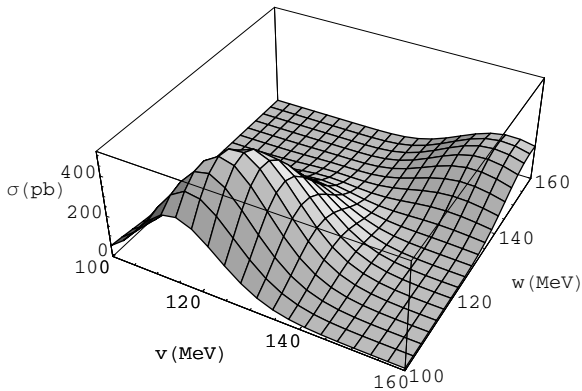}
\end{center}
\caption{The total cross section of $^{12}$C + $^{12}$C $\rightarrow$ $^{24}$Mg + $\pi^{0}$ as a function of the model
parameters $v$ and $w$. The calculation is done for a pion
momentum  $41$ MeV/c which
corresponds to the total energy of about $6$ MeV above  threshold.
\label{wvplot}}
\end{figure}

Here we apply our approach to the cross section of the $^{12}$C + $^{12}$C
$\rightarrow$ $^{24}$Mg + $\pi^{0}$ reaction.  This process, along with its
isospin analog $^{12}$C + $^{12}$C $\rightarrow$ $^{24}$Na + $\pi^{+}$,
represents those few heavy ion pionic fusion reactions for which experimental
data exist \cite{horn96}. The application of the developed theory does not
present a great difficulty except the fact that the cross section is quite
dependent on the structure of initial and final states of interacting
nuclei. Within the harmonic oscillator picture we have approximately $3\times
10^8$ different combinations of interacting states that correspond to the same
energy. Angular momentum and isospin conservation constraints reduce this
number by several orders of magnitude. Additional shell model interactions have
to be introduced to build up a realistic nuclear state for each of the nuclei
and reduce this large number of states, that are degenerate in our model, to
the ones of interest.  Based on this argument we will present here the
Monte-Carlo averaged cross section, where we average over random Cartesian
states.  In the following Fig. \ref{sigma} we display the total reaction cross
section as a function of pion momentum.  We use here the oscillator parameters
$v=104$ MeV/c and $w=119$ MeV/c which are estimated by various theoretical models
\cite{steven1,steven2}.
\begin{figure}
\begin{center}
\epsfxsize=8.0cm \epsfbox{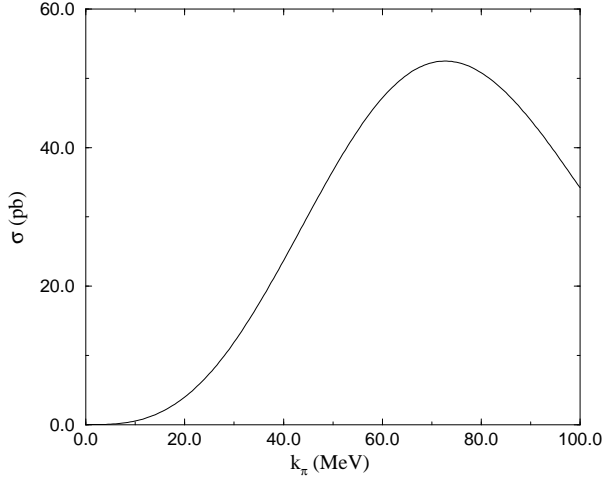}
\end{center}
\caption{The reaction cross section for $^{12}$C + $^{12}$C $\rightarrow$
 $^{24}$Mg + $\pi^{0}$
with
oscillator parameters $v=104$  MeV/c and $w=119$ MeV/c as a
function of pion momentum.  
\label{sigma}}
\end{figure}
The experimentally estimated
cross section for this reaction is  $208\pm 38$ pb
which was observed for
pion momentum  41 MeV/c \cite{horn96}.
In this example we again underestimate the cross section. To see the
sensitivity of our results  
we present in Fig. \ref{wvplot} the dependence of the cross section on
oscillator parameters for pion energy at about $6$  MeV (momentum 41 MeV/c).
This figure indicates that a reasonable variation of parameters could cause a
change in the answer by an order of magnitude.  We emphasize again that in our
calculations we did not project the participating nuclei onto appropriate
shell-model states. Such a projection would require additional nuclear
structure input. Given that the existing experimental data do not clearly
resolve the structure of the final state this seems sufficient.  As a
conclusion, within all the limitations discussed above, the agreement between
the introduced theory and the experimental results of this rare process seems to
be remarkable.

\subsection{Calculations  for heavy nuclei}
In this section we apply the low-momentum approximation for the cross section
described by Eq. (\ref{generic}) to several reactions, with the goal of
understanding the general dependence with respect to the mass of the incoming
nuclei. In order to calculate the cross section, one needs the harmonic
oscillator parameter $v$ which can be estimated from the experimentally
determined r.m.s. radii of the nuclei\cite{vries},

\begin{eqnarray}
r_{\rm r.m.s.}^2={1\over A}\,\sum_i \left< r_i^2 \right> =
= {1\over A}\,\sum_i \frac{1}{v^2}\left( \alpha_i+\frac{3}{2}\right).
\end{eqnarray}

\begin{figure}
\begin{center}
\epsfxsize=8.0cm \epsfbox{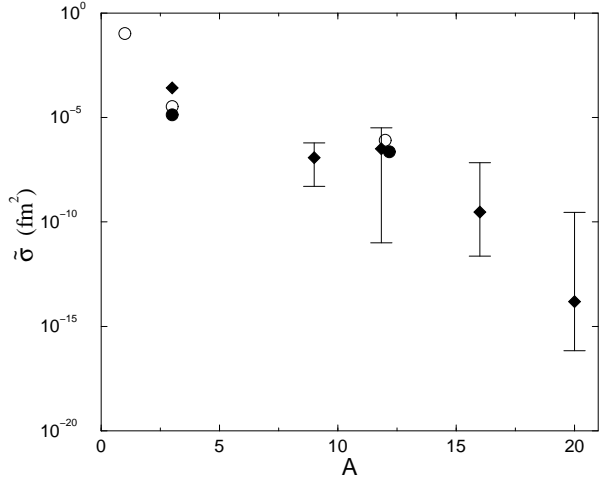}
\end{center}
\caption{The general behavior of the pionic fusion cross section
$A+A\,\rightarrow 2A + \pi$ versus the mass number of initial nucleus $A$. The
plotted value $\tilde{\sigma}$ is related to a total cross section as
$\sigma=\tilde{\sigma}\,( k/m_{\pi})^3$. Calculations in the low-momentum
limit (filled diamonds) show that cross sections fall by several orders of
magnitude in this mass range, but remain in the picobarn region for nuclei as
large as oxygen. The highest and the lowest cross section found
within the shell model configurations are represented by error bars.
Experimental measurements are displayed (open circles) and
compared to calculations (filled circles) which were performed for the finite
pion momenta corresponding to the experiments.
\label{mass_fig}}
\end{figure}

In order to calculate the cross section, one needs to know the incoming energy
of the nuclei as well as the energy of the outgoing pion. Calculations of the
cross sections were performed for incoming nuclei $^9$Be, $^{12}$C,
$^{16}$O and $^{20}$Ne with corresponding fusion products $^{18}$O,
$^{24}$Mg, $^{32}$S and $^{40}$Ca in
the limit of low pion momentum. In this limit the cross section is proportional
to the cube of the pion momentum,
\begin{equation}
\sigma = \tilde{\sigma}(k^3/m_{\pi}^3).
\end{equation}
Values of $\tilde{\sigma}$ are displayed as a function of the mass number of the
incoming nuclei in Figure \ref{mass_fig}. The shell model configurations are
again randomly chosen from the available set of Cartesian states that conserve
isospin and parity. Average values are represented by filled diamonds while the
states with the highest and lowest cross sections are represented by the
boundaries of the error bars. The large error bars demonstrate the wide
fluctuation in strengths for individual states. However, despite the
fluctuations, it is clear that the overall trend is of a decreasing cross
section with increasing mass.

Also shown in Figure \ref{mass_fig} are experimental measurements 
represented by open circles for the $pp$, $^3$He$^3$He and
$^{12}$C$^{12}$C cases
discussed previously. The corresponding calculations, which were
performed for the experimentally measured pion momenta rather than 
in the low-momentum limit are also displayed with closed circles.
One sees that
the cross sections fall by several orders of magnitude, but the
measurements are still feasible 
throughout the wide range of masses.  Calculations could be
performed for heavier nuclei, but for larger masses the Coulomb barrier becomes
important, and shuts off the possibility of fusion for masses larger than 20.

\section{Conclusions}
Near threshold meson production represents a unique area of heavy ion
reactions. In this area the reactions underline the pronounced features of
quantum many-body physics. Most theoretical approaches to understanding and
predicting these phenomena lose their validity in such an extreme regime. In
this paper we have proposed a simple model to study the processes of deep
subthreshold pion production. The pionic fusion cross section was obtained in a
Born approximation with respect to  pion production and in the
sudden approximation for the nuclear rearrangement.
The participating nuclei were described by the harmonic
oscillator shell model in moving oscillator potentials.  The advantage of the
method is that it allows one to incorporate energy, momentum, spin and isospin
conservation laws precisely and respect the Pauli principle at all steps of the
calculation.  Further aspects of nuclear structure could be additionally taken
into account.  At threshold energies these constraints pose the most powerful
restriction on the reaction and cannot be ignored as is done in statistical and
kinetic models which are reasonable at higher energy.
The obvious disadvantage of
the model is that the sudden approximation  does not consider
the slow changes of
the nuclear mean field in the process of interaction.  For the future
developments it seems feasible to incorporate the time dependence and solve the
equations for the evolution of a nuclear mean field parameters. 

The nearly analytical form of overlaps greatly simplified the calculations for
this study.  We used a spherically symmetric nuclear mean field but in some
cases this symmetry prohibits the transition and this would require a
consideration of deformations, i.e. different oscillator parameters for
different directions.  The above mentioned limitations are reflected by the
difficulty in determining the parameters of the model, and lead to about an
order of magnitude ambiguity in the result for $s$-$d$ shell nuclei.  More
realistic single-particle wave functions could be incorporated into the model.
Some of the exponential factors in Eqs. (\ref{ppamp}) and (\ref{p0amp}) arise
directly from the Fourier transformation of the Gaussian tails in harmonic
oscillator wave functions and could be substituted with modified factors that
would reflect a more realistic behavior.

We would like to stress here again that pionic fusion is a very rare process
presenting a tiny fraction of the total cross section.  The agreement that was
observed between  calculations and experimental data for the cross
sections ranging from $10^{-4}$ to $10^{-9}$ barns is remarkable.  Within the
limits of the low pion momentum approximation in the class of the reactions
$A+A \rightarrow 2A +\pi$, we were able to obtain a general formula, Eq.
(\ref{generic}), for the cross sections. The proposed techniques can certainly
be applied in the same manner to other pion production reactions.  The
processes of electrofission \cite{fission} present an another interesting
avenue to exercise this technique.

\acknowledgments{We thank V. F. Dmitriev for constructive
discussions. This work was supported by the National Science Foundation,
grants no 95-12831 and 96-05207.}

\section*{Appendix}
\subsection{Harmonic oscillator wave functions and overlaps
\label{apx_1}}
Work with harmonic oscillator wave functions often involves integrations of the
expressions constructed of polynomials and Gaussian weights. Thus,
the following integral is useful $(a>0)$,
\begin{equation}
\int_{\infty}^{+\infty} x^n e^{-ax^2+ibx}\,dx=\sqrt{\pi\over a} e^{-b^2/ 4a}
\quad a^{-n/2} \,
T_n\left ({i b \over {2 \sqrt a}}\right ) \quad ,
\label{gaussian}
\end{equation}
where $T_n(x)$ is a sum arising from the binomial expansion,
\begin{equation}
T_n(x)=\sum_{j=0,2,4 \ldots }^n {n!(j-1)!! \over j! (n-j)! 2^{j/2}}
x^{n-j} \quad .
\label{T_n}
\end{equation}
This expression can be used for the evaluation of any integral encountered in
this work.  There are two important limiting cases for the sum $T_n(x)$,
$x\rightarrow 0$ and $x\gg 1$:
\begin{equation}
\lim_{x\rightarrow 0} T_n(x)=\left \{ \matrix { {(n-1)!!\over 2^{n/2}} \quad {\rm if} \; n \; {\rm is \; even} \cr
0 \quad {\rm if}\;  n \; {\rm is \; odd} }\right. \quad ,\quad
\lim_{x \rightarrow \infty} T_n(x)= 2^{-n/2} \, x^n \quad .
\end{equation}
The Gaussian-Fourier
integration of Eq. (\ref{gaussian}) is a transformation on the space of
polynomials 
defined on the basis
\begin{equation}
x^n \rightarrow [x^n](p)=T_n(p)
\label{transform}
\end{equation}
The following two-dimensional integrals  often appear in our calculations,
\begin{equation}
\int\int (x-y)^n e^{-a(x^2+y^2)} dx\, dy=\left \{ \matrix{ \pi
a^{-1-n/2} (n-1)!! \quad {\rm if} \; n \; {\rm is \; even} \cr
0 \quad {\rm if}\;  n \; {\rm is \; odd} }\right. \quad ,
\label{oddeven}
\end{equation}
\begin{equation}
\int\int (x-y)^n e^{-a(x^2+y^2)} e^{ip(x-y)} dx\, dy={\pi\over a}
\left ({2\over a} \right )^{n/2} e^{-p^2/2a} T_n\left ({ip\over \sqrt
{2a}} \right ) \quad .
\end{equation}

The basic block of the calculations is the overlap of two one-dimensional
harmonic oscillator wave functions with different oscillator parameters,
shifted locations of the centers and possible additional factor $e^{-i kx}$
that enters the single-particle interaction integral from Eq.(\ref{eikx}).
This type of integral, the generalized Debye-Waller factor,
can be written in a factorized form:
\begin{eqnarray}
\nonumber
\langle (\beta;r)_v|e^{-i k x}|(\alpha;r^{\,\prime})_w\rangle 
=\eta^{1/2}\times
\\
\exp \left({\frac{-k^2} {2(v^2+w^2)}
-\frac{(r-r^{\,\prime})^2v^2w^2} {2(v^2+w^2)}-{\frac{i k(rv^2+r^{\,\prime}w^2)}
{(v^2+w^2)}}} \right )
P_{\beta \, \alpha}((r^{\,\prime}-r),k;v,w) \, ,
\label{SP}
\end{eqnarray}
where $\eta$ is given by Eq. (\ref{eta}) and $P_{n\,m}(r,k;v,w)$ is a dimensionless polynomial of $r$ and
$k$ of the highest power 
$n+m$ with coefficients dependent on $w$ and $v$.
The following are examples of these polynomials for the
smallest values of $n$ and $m$:
\begin{eqnarray}
\nonumber
P_{0\,0}(r,k;v,w)=1 \, , \\
\nonumber
P_{0\,1}(r,k;v,w)=P_{1\,0}(-r,k;w,v)=-{\frac{{\sqrt{2}}\,\left( i\,k + r\,{v^2} \right)
\,w}{{v^2} + {w^2}}} \, ,\\
\nonumber
P_{1\,1}(r,k;v,w)={\frac{2\,v\,w\,\left( {v^2} + {w^2} - \left( k - i\,r\,{v^2} \right) \,
        \left( k + i\,r\,{w^2} \right)  \right) }{{{\left( {v^2} +
{w^2} \right) }^2}}} \, .
\end{eqnarray}
The technique of obtaining these expressions is simple though tedious.  An
important situation $k=0$ would correspond to the overlap of two wave functions
without a pion production, in this case we will not write $k$ as an argument.
It can be shown that \cite{baz}
\begin{equation}
\nonumber
P_{i\,j}(r;v,w)=\sum_{k+l=0,2,4\ldots}
      ^{k=i,l=j}{\sqrt{{\frac{i!\,j!}{k!\,l!}}}}\, 
      {\frac{{{\left( -1 \right) }^{j-l}}\,{v^{j-l}}\,{w^{i-k}}}
        {\left( i-k \right) !\,\left( j-l \right) !}}\,
      {{\left( {\frac{r\,\eta}{\sqrt{2}}} \right) }^
        {i+j-k-l}} P_{k\,l}(0;v,w)
\end{equation}
and
\begin{equation}
\nonumber
P_{k\,l}(0;v,w)=\sqrt{k!\over l!}
{\cal P}_{{(k+l)/ 2}}^{{(l-k)/ 2}}\left (\eta\right
) \quad ,
\end{equation}
with  ${\cal P}_{\alpha}^{\beta}$ being the associated
Legendre polynomials.
A bit simpler case is
\begin{equation}
P_{i\,j}(r;v=w=1)= \sqrt{{i!j!\over
2^{i+j}}} (-1)^j \sum_{k=0}^{min(i,j)} (-1)^k r^{i+j-2k} {2^k \over
k!(i-k)!(j-k)!} \quad .
\end{equation}
Any three-dimensional overlap is reduced to the one-dimensional form of
Eq. (\ref{SP}) in a direct way
\begin{equation}
\langle ({\vec \beta};{\bf
R})|({\vec\alpha};{\bf
r})\rangle =\prod_{x=1,2,3}\langle (\beta_x;R_x)|(\alpha_x;r_x)\rangle
\quad . 
\end{equation}
Similarly we  introduce
\begin{eqnarray}
\nonumber
\langle ({\vec \beta},{\bf r})|e^{-i {\bf k} \cdot  {\bf x}}|({\vec
\alpha},{\bf r}^{\,\prime})\rangle 
=\eta^{3/2} \times
\\
\exp \left ({\frac{-k^2} {2(v^2+w^2)}
-\frac{({\bf r}-{\bf r}^{\,\prime})^2v^2w^2} {2(v^2+w^2)}-\frac{-i{\bf k}({\bf r}v^2+{\bf r}^{\,\prime}w^2)}
{(v^2+w^2)}} \right )
P_{{\vec \beta}\,{\vec \alpha}}(({\bf r}^{\,\prime}-{\bf
r}),{\bf k};v,w) \quad ,
\label{3dSP}
\end{eqnarray}
where
\begin{equation}
P_{{\vec \beta}\,{\vec \alpha}}(({\bf r}^{\,\prime}-{\bf r}),{\bf
k};v,w)=\prod_{x=1,2,3}P_{\beta_x\,
\alpha_x}(r^{\,\prime}_x-r_x,k_x;v,w) \quad . 
\end{equation}
\subsection{Calculational details of the $A+A \rightarrow 2A
+\pi^{+}$ reaction.\label{apx_2}}
Overlaps of many-body nucleon wave functions can be expressed in our
approximation by a  determinant of  single-particle overlaps:
\begin{equation}
\langle ({\vec \beta}_1,\ldots,{\vec \beta}_n;{\bf R})|({\vec\alpha}_1,\ldots;{\bf
r})(\ldots;{\bf r}^{\,\prime})\rangle 
\,=\left|\matrix{
\langle ({\vec \beta}_1;{\bf R})|({\vec\alpha}_1;{\bf r})\rangle  & \cdots &
\langle ({\vec \beta}_1;{\bf R})|({\vec\alpha}_n;{\bf r}^{\,\prime})\rangle   \cr
\vdots & \ddots & \vdots \cr
\langle ({\vec \beta}_n;{\bf R})|({\vec\alpha}_1;{\bf r})\rangle  & \cdots &
\langle ({\vec \beta}_n;{\bf R})|({\vec\alpha}_n;{\bf r}^{\,\prime})\rangle  \cr}
\right |\, .
\end{equation}

Eq. (\ref{3dSP}) allows one to take identical exponential
multipliers in each row
outside the determinant as a common factor in all calculations 
leaving  only the matrix of polynomials
$P_{{\vec \beta}\,{\vec \alpha}}$ to be evaluated.
A simple example of this is the  calculation of
the normalization:
\begin{equation}
\langle ( {\vec \alpha}_1,{\vec \alpha}_2 \ldots {\vec \alpha}_A ;{\bf r})_v)|(
{\vec \alpha}_1,{\vec \alpha}_2 \ldots {\vec \alpha}_A ;{\bf
r}^{\,\prime})_v\rangle  = ||P ({\bf r}-{\bf
r}^{\,\prime}; v,v)|| e^{-A({\bf r}-{\bf r}^{\,\prime})^2 v^2/4}\, e^{i {\bf p}
({\bf r}-{\bf r}^{\,\prime})} \, .
\label{norm_matrix}
\end{equation}
In this expression $||P ({\bf r}-{\bf
r}^{\,\prime}; v,v)||$ is a determinant of a matrix with the entries $P_{{\vec
\alpha}_i \,{\vec \alpha}_j}$. As discussed in Sect \ref{sec_2}, this overlap is
equal to that of the CM wave functions of two harmonic oscillators located at
${\bf r}$ and ${\bf r}^{\,\prime}$. For a nucleon system in the lowest state
(in terms of harmonic oscillator shell excitations), the CM wave function is
the harmonic oscillator wave function of the ground state $|(0,{\bf
r})_{\varsigma}\rangle $. We obtain an interesting mathematical fact
\begin{equation}
||P ({\bf r}-{\bf
r}^{\,\prime}; v,v)||=P_{0\,0}({\bf r}-{\bf r}^{\,\prime}; \varsigma,\varsigma)=1
\, .
\end{equation}
Comparison of the exponents in Eq. (\ref{norm_matrix}) and
Eq. (\ref{3dSP}) gives the value of the oscillator parameter for the
center-of-mass  oscillation as $\varsigma= v \sqrt{A}$.

With the same strategy, one can approach the calculation of the reaction $A+A
\rightarrow 2A+\pi^{+}$ extracting all exponential factors. Corresponding
values of the overlaps $F_n$,$G_l$ and $H_{n\,l}$ may be rewritten, defining
new polynomials $f_n$, $g_l$ and $h_{n\,l}$:
\begin{eqnarray}
F_n=\eta^{3(2Z-1)/2}\times \left \{ \matrix{
\exp\left (-{\eta v w}\left [Z({\bf R}-{\bf r})^2+(Z-1)({\bf R}-{\bf
r}^{\,\prime})^2\right ]/4\right)\,f_n & n\leq Z \, , \cr
\exp\left (-{\eta v w}\left [(Z-1)({\bf R}-{\bf r})^2+Z({\bf R}-{\bf
r}^{\,\prime})^2\right ]/4 \right ) \,f_n & n> Z
\, ;} \right.
\label{fgh}
\\
\nonumber
G_l=\eta^{3(2N)/2}
\exp\left ({-N{\eta v w }\left [({\bf R}-{\bf r})^2+({\bf R}-{\bf
r}^{\,\prime})^2\right ]/4}\right ) \, g_l \, ,
\\
\nonumber
H_{n l}=\eta^{3/2}
\exp \left (-{({{\bf R}-{\bf r}})^2 v^2w^2 
+{\bf k}^2+2 i {\bf k} \cdot ({\bf R}v^2+{\bf r}w^2)
\over 2(v^2+w^2)}\right ) \, h_{n l} \qquad ({\rm if}\, n> Z  ,\,r
\Leftrightarrow r^{\,\prime}) \, .
\end{eqnarray}

It is useful to notice here that all the polynomials are functions of
distances between the nuclei  $({\bf
r}-{\bf R})$ and $({\bf r}^{\,\prime}-{\bf R})$ that we will denote as
${\bf x}$ and ${\bf y}$ respectively. Considering integration in Eq. (\ref{sum})
over variables ${\bf x}$,  ${\bf y}$ and ${\bf R}$ we observe from
Eq. (\ref{fgh}) that the oscillating phase has  the form
\begin{eqnarray*}
\exp\left ({-i {\bf k} \cdot ({\bf R}v^2+{\bf r}w^2)
\over (v^2+w^2)}+i{\bf p}({\bf r}-{\bf r}^{\,\prime})-i{\bf p}_f\cdot
{\bf R}\right )\\
=\exp\left ({i{\bf p}\cdot ({\bf x}-{\bf y})-{i {\bf k}\cdot {\bf x}
w^2 \over v^2+w^2} - i ({\bf k}+{\bf p}_f) \cdot {\bf R}}\right ) \, ,
\end{eqnarray*}
and integration over ${\bf R}$ gives a momentum preserving
$\delta$-function that requires  ${\bf k}=-{\bf p}_f$.
For convenience we split the sum in Eq. (\ref{sum}) over $n\leq Z$ and $Z<n\leq
2Z$ and substitute $F$, $G$ and $H$ from Eq. (\ref{fgh})
\begin{eqnarray}
\nonumber
\langle F|H|I\rangle ={1\over {\cal N}_i {\cal N}_f}\int{V\over{\sqrt{2\omega V}}} {1\over 2m}
\eta^{3A}\,e^{-Avw\eta(x^2+y^2)/4}\,
e^{-k^2/2(v^2+w^2)}
\\
=\left (e^{-i {\bf k} \cdot {\bf x} \eta w/2v}
\sum_{i\leq Z, \, j} (-1)^{i+j}f_i g_j h_{ij} 
+ 
e^{-i {\bf k}\cdot {\bf  y} \eta w/2v}\sum_{i>Z, \, j} (-1)^{i+j} f_i g_j h_{ij}
 \right ) e^{-i{\bf p} \cdot ({\bf x}-{\bf y})} d^3x d^3y\, .
\label{ampl}
\end{eqnarray}
The terms $\sum f_i g_j h_{ij}$ are again some polynomials of ${\bf x}$ and
${\bf y}$ proportional to $|k|$ and containing parameters $v$ and $w$.  The
final integration can be performed with the help of Eq. (\ref{gaussian})
corresponding parameters $a$ and $b$ being
\begin{equation}
a=A\eta v w /4 \quad , \quad b=\pm\, p-k\eta w/2v \quad .
\end{equation}
As a result,  we arrive at the formula  (\ref{ppamp})
with polynomials
\begin{eqnarray}
\nonumber
P(k,p)={1\over |{\bf k}|}\left [\sum_{i\leq Z,\, j} (-1)^{i+j}f_i g_j h_{ij}\right ]
\left (-i\,{{\bf p}+{\bf
k}\eta w/2v \over\sqrt{A \eta v w }} \, ,\,
{i{\bf p}\over\sqrt{A \eta v w }} \right )\, ,\\
Q(k,p)={1\over |{\bf k}|} \left [\sum_{i>Z,\, j} (-1)^{i+j}f_i g_j h_{ij}\right ]
\left ({-i{\bf p}\over \sqrt{A \eta v w }} \, , \,
i\,{ {\bf p}-{\bf
k}\eta w /2 v \over\sqrt{A \eta v w }}
\right ) \, ,
\label{PQ}
\end{eqnarray}
where the first argument is the transformation of elements of vector ${\bf x}$
and the second that of vector ${\bf y}$. From here it is also seen that if
before transformation there existed a symmetry between ${\bf x}$ and ${\bf
y}$, i.e. the nuclei were in an identical state, then $P({\bf k},-{\bf p})=Q({\bf k},{\bf
p})$.

\subsection{Toward a complete analytical answer, reaction $A+A\rightarrow 2A +
\pi^{0}$. \label{apx_3} }

As it was pointed out in the main text, the amplitude of the pionic process
is approximately proportional to the amplitude of the fusion reaction. One can
study the properties of the determinants arising in a fusion reaction in a
quite general way, separately considering the four types of particles
distinguished by spin and isospin in the reaction of fusion of the type $A+A
\rightarrow 2A $.  This leads to the following form of a single-particle
overlap matrix
\begin{equation}
2A \left \{ \left ( \matrix{
\langle ({\vec \beta}_1;{\bf R})|({\vec\alpha}_1;{\bf r})\rangle  \cdots &  & \cdots
\langle ({\vec \beta}_1;{\bf R})|({\vec\alpha}_A;{\bf r}^{\,\prime})\rangle   \cr
\vdots &  & \vdots \cr
\underbrace{\langle ({\vec \beta}_A;{\bf R})|({\vec\alpha}_1;{\bf r})\rangle 
\cdots}_{{\rm first}\, A \, {\rm nucleons}} & & \underbrace{\cdots
\langle ({\vec \beta}_A;{\bf R})|({\vec\alpha}_A;{\bf r}^{\,\prime})\rangle }_{{\rm second}
\, A \, {\rm nucleons}}\cr}
\right ) \right . \, .
\end{equation}

Without loss of generality, ${\bf R}$ can be set to zero. A second important
feature is that in nuclei under consideration all inner shells are
filled. Therefore, the resulting determinant is a function of a nucleon number
$A$ and extra parameters arising from different ways to distribute the
particles in the outer shells.

It is interesting to present the exact result for the one-dimensional case
where the problem is uniquely defined. We consider two oscillators with
single-particle states from $0$ till $A-1$ overlapping with one larger
oscillator with occupied states from $0$ up to $2A-1$,
\begin{eqnarray}
\nonumber
 \left |\matrix{
\langle (0;0)_v|(0;x)_w\rangle  & \cdots & \langle (0;0)_v|(A-1;y)_w\rangle  \cr
\vdots & \ddots & \vdots \cr
\langle (2A-1;0)_v|(0;x)_w\rangle  & \cdots & \langle (2A-1;0)_v|(A-1;y)_w\rangle  \cr}
\right |
\\={\frac{{(-1)^A (w(x-y))^{A^2}\sqrt{\left
          ( 2\,A \right) !}}}{{2^{{\frac{\left(A-1 \right) \,A}{2}}}}\,{\sqrt{A!
}}\,
     \left( \prod_{i = 1}^{A}{\frac{\left( 2\,i \right) !}{i!}}
\right) }}\, \eta^{(2A-1)A}={1\over
\gamma} (w(x-y))^q \,\eta^{\cal Q}
 \quad ,
\label{onedim}
\end{eqnarray}
\begin{equation}
\gamma (A)={(-1)^A\,\sqrt{A!} \over \sqrt{(2A)!}}{{2^{{\left(A-1
     \right) \,A}/{2}}}\,
     \left( \prod_{i = 1}^{A}{\frac{\left( 2\,i \right) !}{i!}}
\right) }\, .
\end{equation}
The result is just a single term which depends only on the distance
between the two
initial oscillator locations raised to the power equal to the difference in
total number of quanta between initial and final systems, $q=A^2$. The term
$\eta=2vw/(v^2+w^2)$ comes in the power of total number of quanta in the final
nucleus, ${\cal Q}=(2A-1)A$. This remains true only for Fermi systems in the
ground state, i.e. if there are no gaps in the harmonic oscillator
single-particle level occupation.  The situation for a three-dimensional
oscillator is similar.  The required polynomial is still given by one term that
has a form of the product
\begin{equation}
{1\over \gamma} (x_x-y_x)^{q_x} (x_y-y_y)^{q_y} (x_z-y_z)^{q_z} w^{q_x+q_y+q_z}
\,\eta^{{\cal Q}_f} \quad ,
\label{polinom}
\end{equation}
where integers  $q_x$, $q_y$ and $q_z$ are  differences of the number of
quanta between the final and initial systems in $x$, $y$ and $z$ directions,
respectively. 
A specific three-dimensional  complication arises from the
following aspect. The lowest energy state is, in general, degenerate
as for non-magic nuclei one has the freedom of placing several
particles into $(n+1)(n+2)/2$ degenerate levels of the $n$-th
shell. The numerical parameter $\gamma$ depends in this case  also 
on the way the particles are placed in the outer shell of each
nucleus. 
The harmonic oscillator symmetries in the problem  often prohibit the
transition.

The polynomials in
Eq. (\ref{ampl})  acquire a form of a product of four components,
each of the form of Eq. (\ref{polinom}) for each type of nucleons,
times the sum of terms $({\vec \sigma} \cdot {\bf k})$ acting on
every pair of interacting nucleon species. 
Using the integrals from Eq. (\ref{oddeven}) and writing the action of
$({\vec \sigma} \cdot {\bf k})$ between initial and final spin parts
of the wave function as a matrix element ${\tilde M}$
we arrive at
the  expression for the polynomial  in Eq. (\ref{p0mat}) 
\begin{eqnarray}
\nonumber
P_{q_x,q_y,q_z}(k=0,p)={1\over \gamma} \left ({2 (v^2+w^2) \over A v^2} \right
)^{(q_x+q_y+q_z)/2}\, \eta^{{\cal Q}_f} \times
\\
\left (
(q_x-1)!! \, (q_y-1)!! \, 2^{q_z\over 2} T_{q_z}\left (
{ip\sqrt{2\over A \eta v w }} \right ) \right ){\tilde M} \quad .
\label{det}
\end{eqnarray}
In the above expression we have redefined $\gamma$ as a product of
$\gamma$'s for all four types of nucleons.

\end{document}